\documentstyle[aps,epsf]{revtex}
\newcommand{\be}{\begin{equation}}
\newcommand{\ee}{\end{equation}}
\newcommand{\Leq}[1]{\label{eq:#1}}
\newcommand{\Req}[1]{\ref{eq:#1}}
\newcommand{\bea}{\begin{eqnarray}}
\newcommand{\eea}{\end{eqnarray}}
\begin{document}
\begin{center}
{\bf QUANTUM BREAKING TIME SCALING IN THE SUPERDIFFUSIVE DYNAMICS}
\end{center}

\begin{center}
{\bf A. Iomin$^{(a)}$ and George M. Zaslavsky$^{(b)}$ \\
Courant Institute of Mathematical Sciences,
New York University, \\
251 Mercer St., New York, NY 10012, USA}
\end{center}

\begin{abstract}
We show that the breaking time of quantum-classical correspondence
depends on
the type of kinetics and the dominant origin of stickiness. For sticky
dynamics of quantum kicked rotor, when the hierarchical set of islands
corresponds to the accelerator mode, we demonstrate by simulation that the
breaking time scales as $\tau_{\hbar} \sim (1/\hbar )^{1/\mu}$
with the transport exponent $\mu > 1$ that corresponds to superdiffusive
dynamics \cite{sz}. We discuss also other possibilities for the breaking
time scaling and transition to the logarithmic one 
$\tau_{\hbar} \sim \ln(1/\hbar )$ with respect to $\hbar$.

\end{abstract}

\vspace{3ex}

1. Classical chaotic dynamics can be characterized by a Lyapunov
exponent
$\Lambda$ and infinitely divisible filamentation of the phase flux.
Quantized
procedure stops the filamentation due to the uncertainty principle
and,
as a
result, breaks the applicability of semiclassical approximaiton. The
corresponding breaking time was found in \cite{bz}
\be
\tau_{\hbar} = (1/\Lambda ) \ln (I_0 /\hbar ), 
\Leq{e1}
\ee
where $I_0$ is a characteristic action, indicating a fast (exponential)
growth of quantum corrections to the classical dynamics due to chaos.
The origin of this time was explained in detail in \cite{zasl}  and gained
a wide discussion \cite{3z}. The logarithmic scaling in $\hbar$ for
$\tau_{\hbar}$
corresponds to a fairly good and uniform chaotic mixing (see also a
final comment in \cite{log}).

Typical Hamiltonian chaotic dynamics is not ergodic due to the
presence of infinite islands set in phase space \cite{6z}, and Lyapunov
exponent is 
not uniform due to cantori and possible hierarchical structures of
islands \cite{6z} and their stickiness \cite{kar,7z}. This complicates a
process of diffusion transforming the transport from the Gaussian type to 
the anomalous (fractional) one \cite{hcm,zen}. Particularly, sticky
properties of the island boundaries should impose algebraic laws of the 
survival probability 
\be
P(t) \sim 1/t^{\gamma} 
\Leq{e2}
\ee
of a particle to escape after time $t$ from a domain near the islands
\cite{hcm,7z}. The immediate consequence from the scaling property
(\Req{e2}) is its breaking at some critical time $\tau^*$ for the case of
quantum chaos since a hierarchical dynamical chain has no limit
$t\rightarrow\infty$ and should be abrupted when an island in the
chain reaches area $S^* = \hbar$ and the quantum effects become
important.
This comment, starting from \cite{fgp1}, was discussed in detail in 
\cite{log}, where a power law for the Planck constant scaling was suggested
\be
\tau_{\hbar}^* \sim 1/\hbar^{\delta} 
\Leq{e3}
\ee
for the breaking time of classical considertion applicability with a
value of
$\delta$ depending on the type of the classical algebraic escape time
distribution. There were no quantum simulation in these works and no
definite
values of $\delta$. Discussion of the algebraic law (\Req{e3}) started in
\cite{log,fgp1}, was continued in recent publications
\cite{cms1,savsok,frahm,ket,cgm,ketz2} on the basis of simulations
of quantum maps with an explicit evaluation of (\Req{e2}).

The questions arise: What is the actual scaling of breaking time of
quantum-classical correspondence with respect to the Planck constant, 
logarithmic or algebraic? How universal are values of $\gamma$ or
$\delta$? Using some results of \cite{sz,iz} on the strong
delocalization effects,
and simulation for the quantum kicked rotor (QKR) we will show here that
the result (\Req{e1}), being valid for the cases of normal (Gaussian type)
diffusion,
appears to be an algebraic of the type (\Req{e2}) when diffusion becomes
anomalous, i.e. superdiffusion, with the second moment of the truncated
distribution function as
\be
\langle p^2 \rangle \sim t^{\mu} 
\Leq{e4}
\ee
and transport exponent $\mu > 1$. We were able to obtain
$\tau_{\hbar}^*$
from the simulation as a crossover time of the survival probability 
that  changes the exponent of its algebraic behavior, and compare
values of $\delta$, predicted
by the theory, with the one obtained from the simulation.

2. We consider QKR that corresponds to the standard map in the classical
limit
\be
p^{\prime} = p + K \sin q \ , \ \ \  \
q^{\prime} = q+p^{\prime} 
\Leq{e5}
\ee
defined on the cylinder $p \in (-\infty ,\infty )$,
$q \in (-\pi ,\pi )$ with a control parameter $K$ and the Lyapunov
exponent
$\Lambda \sim 1/\ln K$ for $K \gg 1$ and for almost all domain excluding
areas where $K |\cos q| < 1$. The marginally stable points are defined
by the
conditions $K_m = 2\pi m$, $p = 2\pi n$, $q = \pm \pi /2$ with integers
$(m,n)$. For
\be
0 < K - K_m < \Delta K_m 
\Leq{e6}
\ee
a new set of islands appears \cite{kar,mel,vered}, called tangle
islands in \cite{vered}, as a
result of bifurcation. Dynamics inside the islands is known as the
accelerator mode \cite{8z,9z} and we will call them accelerator mode
islands (AMI). Changing
of $K$ within the interval (\Req{e6}) influences strongly the topological
structure of AMI and consequently the values of the transport
exponent
$\mu = \mu (K)$ \cite{10z,zen}, since the stickiness of trajectories to
the islands boundaries can be different for different island topologies.

It was established in \cite{zen} that for a special
(``magic'') value of $K \equiv K^*$ = 6.908745$\ldots$ the stickiness
can be especially what makes this case to be convenient to
study properties of
the anomalous transport. For the $K^*$ it appears a hierarchical set of
islands-around-islands with the islands sequence $3-8-8-8-\ldots$. The
islands chain satisfies the renormalization conditions
\bea
S^{(n+1)} = \lambda_S S^{(n)} \ , \ \ \ \
T^{(n+1)} = \lambda_T T^{(n)} \ , \nonumber \\
N^{(n+1)} = \lambda_N N^{(n)} \ , 
\Leq{e7}
\eea
where $n$ is a number in the hierarchy sequence, $S^{(n)}$ is an island
area, $T^{(n)}$ is a period of the last invariant curve of the
corresponding
island, $N^{(n)}$ is a number of islands in the chain of the $n$-th
hierarchy level, and $\lambda_S < 1$,
$\lambda_T > 1$,
$\lambda_N > 1$
are some scaling parameters. The renormalization transform (\Req{e7}) can
be
also
extended for Lyapunov exponents $\Lambda$ in a sticky area of the
island's
boundary of the $n$-th level of the island's hiearchy \cite{10z}:
\be
\Lambda^{(n+1)} = \lambda_L \Lambda^{(n)} = \lambda_L^n \Lambda^{(0)} \
,
\ \ \ \ (\lambda_L < 1) \ . 
\Leq{e8}
\ee

In the absence of the island's hierarchy, we get just the result 
(\Req{e1}) with
$\Lambda = \Lambda^{(0)}$. In the presence of the island's hierarchy we
can introduce particle  flux $\Phi^{(n)}$ in phase space through the
island's chain of the $n$-th hierarchical level. It reads
\bea
\Phi^{(n)} = S^{(n)} N^{(n)} /T^{(n)} =
\Phi^{(0)} (\lambda_S \lambda_N /\lambda_T )^n \ , \nonumber \\
\Phi^{(0)} =S^{(0)} N^{(0)} /T^{(0)}   
\Leq{e9}
\eea
in correspondence to (\Req{e7}) and (\Req{e8}). For $K=K^*$ and the
corresponding island's
hierarchy $\lambda_N = \lambda_T$ \cite{zen} and thus
\be
\Phi^{(n)} = \lambda_S^n \Phi^{(0)} \ . 
\Leq{e10}
\ee

The quantum mechanical consideration of the proliferation of islands is
meaningful until the smallest island size is of the order of $\hbar$.
Therefore we get from (\Req{e7}),(\Req{e9}),(\Req{e10}):
\bea
S_{\min} = \hbar = S^{(n_0 )} = S^{(0)} \lambda_S^{n_0} \ ,
\nonumber \\
\Phi_{\min} = \Phi^{(0)} \lambda_S^{n_0} = \lambda_S^{n_0} S^{(0)}
N^{(0)} /T^{(0)} = \hbar N^{(0)} /T^{(0)} \ ,
\Leq{e11}
\eea
and the quantum ``cut-off'' of the hierarchy appears at
\be
n_0 = |\ln (\hbar /S^{(0)} )| / |\ln \lambda_S | \equiv
|\ln \tilde{h} | /| \ln \lambda_S | \ ,
\Leq{e12}
\ee
where we introduce a dimensionless semi-classical parameter 
$\tilde{h} = \hbar /I_0$ and specify
$S^{(0)} \equiv I_0$. Particularly, for the hierarchy at $K=K^*$ we have
$N^{(0)} = 3$ and $\lambda_N = 8$ but it could be many other hierarchies
(see more in \cite{11z}). After the substitution $\Lambda = \Lambda^{(n_0 )}$
we get with (\Req{e8}) and (\Req{e12}):
\be
\tau_{\hbar} = (1/\Lambda^{(n_) )} ) \ln (1/\tilde{h} ) =
(1/\tilde{h} )^{1/\mu} \ln (1/\tilde{h} ) 
\Leq{e13}
\ee
with
\be
\mu = |\ln \lambda_S | /\ln \lambda_T \ .
\Leq{e14}
\ee
The expression of $\mu$ through $\lambda_S ,\lambda_T$ was found in
\cite{10z} for the considered islands hierarchy and the expression for
$\tau_{\hbar}$
coincides with the obtained in \cite{sz} in a different way. The
expression (\Req{e13})
is close to (\Req{e3}) up to a logarithmic term and defines
\be
\delta = 1/\mu, 
\Leq{e15}
\ee
which for $K=K^*$ provides $\mu = 1.25$ (see \cite{zen}) and $\delta$ =
0.8.
When
there is no islands hierarchy, we may put $\lambda_S \rightarrow 0$ or
$n_0 \rightarrow 0$. This yields transition from the algebraic law
(\Req{e13}) to the
logarithmic one for the breaking time and resolves the paradox discussed
in \cite{log}. In addition to this, it was shown in \cite{zen} that for
the considered islands hierarchy
\be
\gamma = 1 + \mu = 1+ |\ln \lambda_S | /\ln \lambda_T \ , 
\Leq{e16}
\ee
what gives $\gamma = 2.25$ for $K=K^*$.  Just these values of
$\gamma$ and $\delta$ we have checked by a simulation.

3. Numerical study of the problem is based on investigation of the
quantum survival probability in some domain $\Delta p \in (-\pi ,\pi )$ that
includes the islands hierarchy \cite{iz3}. The main difficulty appears
due to a part of the wave function that belongs to an island interior
and
``flies'' fast along $p$. A simple way to avoid this type of the
``escape'' from $\Delta p$ is to apply a shift operator
\be
\hat{J} = \exp \{-(2\pi/\tilde{h}) \partial /\partial \hat{n}\} = 
\exp (-2\pi iq/\tilde{h}) \ ,    
\Leq{e17}
\ee
where $q$ is a dimensionless coordinate, $\hat{p} = \tilde{h} \hat{n} =
-i\tilde{h} \partial /\partial q$ is a dimensionless momentum operator
and the wave function $\Psi_t$ at a discrete time $t$ is considered
in the coordinate space. In analogy to \cite{cms1,cms2} we also
introduce the absorbing boundary conditions at the edges of the interval
$\Delta n \in [-(N+1)/2,(N-1)/2]$ for the momentum eigenvalues $\tilde{h} 
n$.
Then the quantum map that keeps the information of the trapping into
$\Delta n$ part of the wave function is
\be
\Psi_{t+1} = \hat{\cal P} \hat{J} \hat{U} \Psi_t 
\Leq{e18}
\ee
with the evolution operator
\be
U = \exp \big(-i\tilde{h} \hat{n}^2 /2 \big) 
\exp\big (-i(K/\tilde{h}) \cos q \big) \ .
\Leq{e19}
\ee
As usual, for the numerical convenience the dimensionless Planck
constant $\tilde{h}$ is taken in a form $\tilde{h} = 2\pi /(N+g)$, 
where $g = (\sqrt{5}-1)/2$  is the inverse golden mean.

The survival probability is defined as
\be
P(t) = |\Psi_t |^2 
\Leq{e20}
\ee
together with definition (\Req{e18}). Simulation was performed for 12
values of
$N$ from the interval $(5\cdot 10^3 \div 7.5\cdot 10^4 )$ which corresponds
 to a good semi-classical approximation for a fairly long time until
it 
fails.

A typical behavior of $P(t)$, obtained for $K=K^*$, is shown in Fig. 1.
It consists of the crossover from the classical behavior (\Req{e2}) with
$\gamma\approx 2.25$ (the same as in \cite{zen}) to some very different
dependence.
The crossover points $\tau_{\hbar}$ can be identified in a way as it is
shown in Fig. 1 for different values of $N$, i.e. $\tilde{h}$. The
corresponding result is presented in Fig. 2. It provides the value of
$1/\delta \approx 1.33 \pm 0.36 \approx \mu$ in a good agreement to the
presented theoretical  estimation (see (\Req{e15})).

4. In conclusion, we need to underline that there is no unique
scenario of chaotic diffusion in classical limit and therefore one may
expect
no unique breaking mechanism of quantum classical correspondence.
Stickiness
and algebraic kinetics through cantori was considered in \cite{hcm}
with
a specific choice of the Markov tree that defined a type of kinetics and
corresponding scales. In this consideration a special hierarchical set
of resonance islands was selected, while in \cite{zen,10z} the Markov
tree was constructed for the tangle islands (see more about the 
classification in \cite{vered}). The difference in the choice of the
islands set selection is imposed by the value of $K$. In this article we 
choose $K=K^*$ which leads to (\Req{e15}) while in
\cite{ketz2,ketz1,ketz3}
the value $K \leq 2\pi$ was selected which probably leads to
stickiness phenomenon described in \cite{hcm} and to the value
\be
\delta = 1/\gamma \ .     
\Leq{e21}
\ee
This value was not linked to the transport exponents $\mu$. Let us
mention also the value $\delta\sim 0.5$ proposed in \cite{cms2} for  
$K = 2.5$ when sticky islands set appears without accelerator mode and 
with no superdiffusion. 
\vspace{.3in}

We thank S. Fishman for stimulating discussions. A.I. is grateful to
T. Geisel for his hospitality at MPI f\"{u}r Str\"{o}mungsforschung
where part of the work has been done and A.I. also thanks L. Hufnagel, R.
Ketzmerick, T. Kottos, H. Schanz, and M. Weiss for valuable and stimulating
discussions.
This research is supported by U.S. Department of Navy Grants Nos.
N00014-96-10055 and N00014-97-0426, and by U.S. Department of Energy
Grant No. DE-FG02-92-ER54184. A.I. was also supported by the Minerva Center of
Nonlinear Physics of Complex Systems and by the Israel Science Foundation.

\section*{Figure Captions}

\begin{description}

\item[Fig.~1]
Typical evolution of the quantum survival probability for $ N=25557 $.
The dashed lines, which correspond to the asymptotics slopes,
determine the breaking point $ \tau_{\hbar} $.

\item[Fig.~2]
The quantum breaking points $ \tau_{\hbar} $ vs dimensionless
semiclassical parameter $ \tilde{h} $. The solid line
with the slop $ 1/\delta\approx 1.33\pm 0.36 $ corresponds to least square
calculations, and the dashed line is the analytical prediction
with $ \delta=1/\mu $.

\end{description}


\begin{thebibliography}{99}
\bibitem{3z} See also \cite{shep}-\cite{cis}, \cite{4z}, and references
therein. Expression (\Req{e1}) was proved together with evaluation of the
remainder, in \cite{5z}. In same articles $ \tau_{\hbar} $ is called 
Ehrenfest time to underline that it shows a breaking of the semiclassical
approximation for the Ehrenfest equations. One may consider this confusing
since the original Ehrenfest equations are exact.
\end{thebibliography}
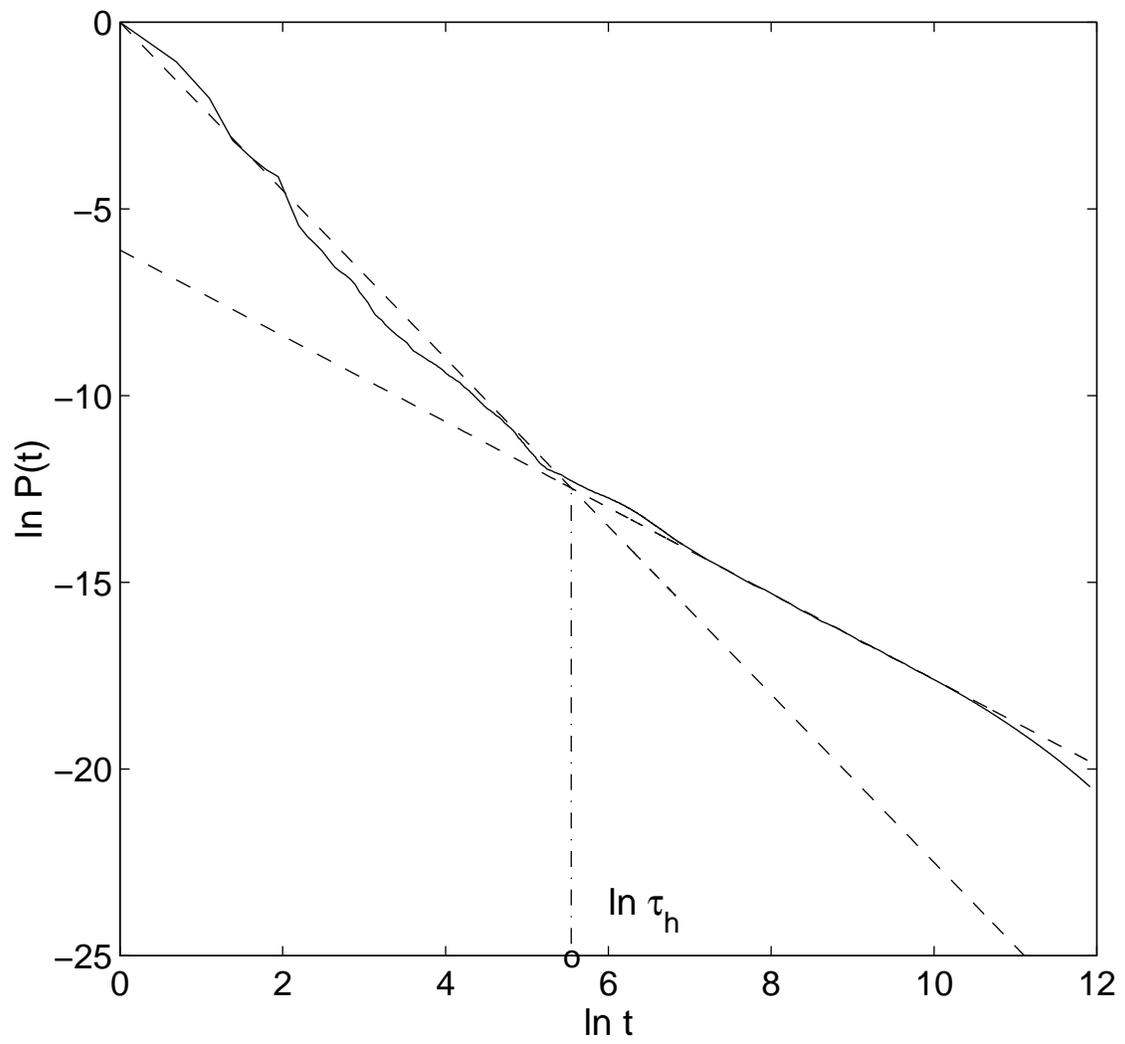
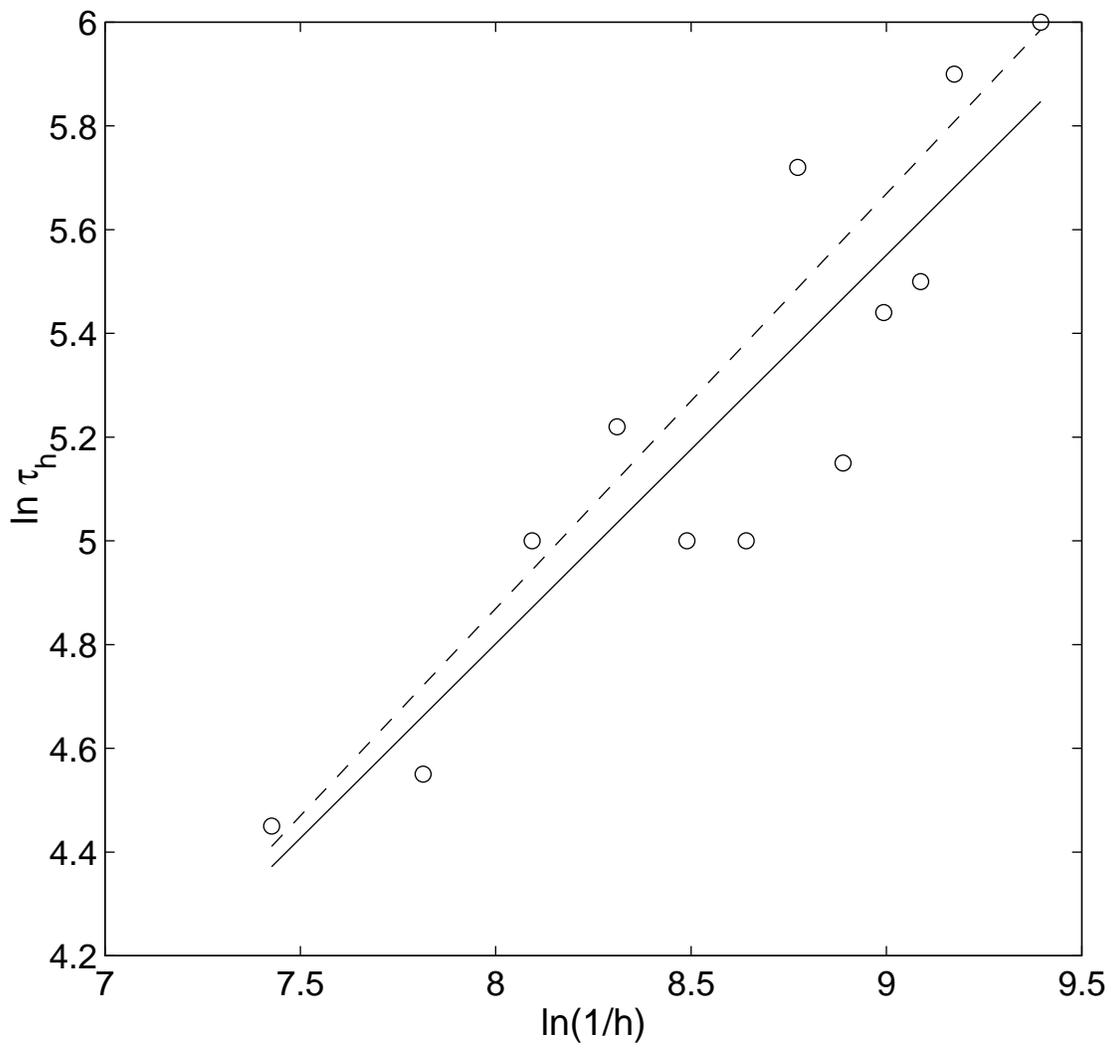
\end{document}